\documentclass[aps,prx,twocolumn,showpacs,floatfix,preprintnumbers]{revtex4-2}
\usepackage{graphics,amssymb,amsmath,epsfig,color,textgreek}
\usepackage{graphicx}
\usepackage[dvipsnames]{xcolor}
\usepackage{dcolumn}
\usepackage{bm}
\usepackage[colorlinks=true,citecolor=cyan,breaklinks=true,pdftex]{hyperref}
\usepackage{breakurl}
\usepackage{subcaption}
\hypersetup{colorlinks=true,citecolor=cyan,linkcolor=red,urlcolor=magenta}
\usepackage{braket}

\begin{document}
\preprint{RIKEN-iTHEMS-Report-24}

\title{Classical optimization with imaginary time block encoding on quantum computers: The MaxCut problem}
\author{Dawei Zhong}
\email{daweiz@usc.edu}
\affiliation{Department of Physics \& Astronomy, University of Southern California, Los Angeles, CA 90089, USA}
\author{Akhil Francis}
\email{afrancis@lbl.gov}
\affiliation{Applied Mathematics and Computational Research Division, Lawrence Berkeley National Laboratory, Berkeley, CA 94720, USA}
\author{Ermal Rrapaj}
\email{ermalrrapaj@lbl.gov}
\affiliation{Lawrence Berkeley National Laboratory, Berkeley, CA, 94720, USA}
\affiliation{Department of Physics, University of California, Berkeley, CA 94720, USA}
\affiliation{RIKEN iTHEMS, Wako, Saitama 351-0198, Japan}

\begin{abstract}
    Optimization problems in finance, physics and computer science are typically very hard to tackle in classical computing and quantum computing could help speed up computations and provide efficient methods for tackling large problems. Typically, to treat the problem with a quantum computer, the optimal solution is cast as the ground state of a diagonal Hamiltonian. We develop a new method, called ITE-BE, based on a recent imaginary time algorithm, which requires no variational parameter optimization as all parameters can be derived analytically from the target Hamiltonian. We also demonstrate that our method can be successfully combined with other quantum algorithms such as quantum approximate optimization algorithm (QAOA). For illustration, here we study the MaxCut problem. We find that the QAOA ansatz increases the post-selection success of ITE-BE, and shallow QAOA circuits, when boosted with ITE-BE, achieve better performance than deeper QAOA circuits. For the special case of the transverse initial state, we adapt our block encoding scheme to allow for a deterministic application of the first layer of the circuit. 
\end{abstract}
\maketitle

\section{Introduction}
The advent of quantum computers has spurred a wide range of research into quantum advantage: problems that can be solved much faster, and/or at a much larger scale through quantum computing. Some of the most well known cases are integer prime factorization \cite{shor1994algorithms}, and unstructured search \cite{grover1997quantum}, to name a few. An interesting venue of research is discrete optimization problems~\cite{abbas2023quantum}. These problems can be expressed as the minimization of a polynomial function of binary variables, and the cost function can be written as a diagonal Hamiltonian which encodes the optimal solution in the ground state~\cite{barahona1982computational,Lucas:2015}.
For instance, Quadratic Unconstrained Binary Optimization (QUBO) models several problems such as MaxCut, minimum vertex cover, and graph coloring~\cite{kochenberger2014unconstrained, mazumder_five_2024}. These problems have found applications in industry~\cite{streif2021beating,shaydulin2024evidence}, condensed matter physics~\cite{Mezard:1986}, and beyond. 

Apart from special cases, finding the exact ground state of such diagonal Hamiltonians is usually difficult, and generally NP-hard \cite{Lucas:2015, barahona1982computational}. Often classical heuristic and approximate methods, such as simulated annealing~\cite{kirkpatrick1983optimization} or semi-definite programming \cite{goemans1995improved}, are used to approximate solutions. Simulated annealing motivated by the annealing process in statistical mechanics, is a probabilistic method where an appropriate temperature and annealing schedule is defined to `cool' the system, while the new configuration is accepted with an appropriate acceptance probability. 

Semi-definite programming (SDP) minimizes a linear cost function subject to a semi-definite constraint, resulting in convex optimization problem which can be solved efficiently \cite{vandenberghe1996semidefinite}. SDP could be applied by relaxing the original optimization problems (e.g., the MaxCut problem), followed by an appropriate rounding scheme to obtain approximate solutions \cite{goemans1995improved}.

Quantum computers could potentially find solutions beyond the realm of classical computation \cite{abbas2023quantum} as quantum hardware moves beyond the noisy intermediate-scale quantum (NISQ) era to fault-tolerant computing \cite{lotshaw2022scaling}. Many quantum algorithms such as quantum approximate optimization algorithm (QAOA) \cite{farhi2014quantum,farhi2016quantum,lloyd2018quantum,farhi2017quantum,Decker:2016}, adiabatic evolution \cite{farhi2001quantum}, or imaginary time evolution methods~\cite{mcardle2019variational,Motta:2020,Huang:2023,Leadbeater:2023} have been proposed to help determine the ground state. All these methods have their own drawbacks too, such as barren plateaus \cite{wang2021noise, holmes2022connecting} and other optimization related issues for QAOA~\cite{bittel2021training}, longer depth circuits for implementing adiabatic evolution\cite{van2001powerful} and larger number of measurements for successful post-selection of block encoded implementation of imaginary time evolution\cite{rrapaj_exact_2024}.

QAOA has been widely explored recently in large sparse graphs, and spin glasses~\cite{basso_spin_glass}, in particular in Sherrington-Kirkpatrick (SK) model  of spin glass \cite{farhi2022quantum}, on MaxCut problems \cite{crooks2018performance} in both software and hardware~\cite{10313920, zhou_quantum_2020, harrigan2021quantum}. In addition to two-body interactions, QAOA has also been applied to four-body interactions, such as the low autocorrelation binary sequences (LABS) problem \cite{shaydulin2024evidence}.

Imaginary time evolution, otherwise known as Wick rotation~\cite{wick}, connects Euclidean and Minkowski space by performing an analytical continuation from real to imaginary time (see figure 6.1 from~\cite{Peskin:1995ev}), and has been used extensively in auxiliary field diffusion Monte Carlo studies~\cite{PEDERIVA:2004,Lonardoni:2018,Lee:2022}. This approach exponentially improves the overlap towards the ground state when applied to an initial state with non-zero overlap with the ground state. Since it is a non-unitary operation, implementing it on a quantum computer is not trivial, and there has been several approaches in the literature. One approach is to find equivalent unitary evolutions, with appropriate quantum measurements and  classical computations, such as in variational imaginary time evolution \cite{mcardle2019variational, gacon2024variational, bauer_combinatorial_2023} where a fixed ansatz is updated to follow McLachlan's variational principle. Another approach discretizes the imaginary time evolution and approximates each step by a unitary constructed through tomography~\cite{Motta:2020,huang2023efficient,turro2022imaginary}. These methods require classical processing and multiple measurements during the time evolution which scale with both system and length of time.
Other methods such as block-encoding schemes \cite{childs2012hamiltonian,low2017optimal,gilyen2019quantum}, encode the desired non-unitary operation in a bigger unitary matrix, which can then be implemented with ancilla qubits and post-selection. Recently an exact block-encoding scheme for discretized imaginary time evolution, based on unitary Restricted Boltzmann Machines (RBM), has been proposed \cite{rrapaj_exact_2024} (for similar recent implementations see~\cite{Leadbeater:2023}).

Here we utilize this imaginary time evolution block encoding (ITE-BE) scheme of the first order Trotter decomposition,  to implement imaginary time evolution in solving classical optimization problems such as MaxCut. We also implement a new parametrization that allows to correct for failed post-selection when the initial state is in the traverse state.  We find that ITE-BE is beneficial in providing solutions to MaxCut, and also as a boosting technique to improve other methods such as QAOA. In Sec.~\ref{sec:prem} we define the MaxCut problem, QAOA and the block encoding method ITE-BE. Then, we proceed to perform numerical evaluations of these methods in Sec.~\ref{sec:eval}. In Sec.~\ref{sec:dis} we summarize our findings and discuss their implications.

\begin{figure*}[!ht]
     \centering
     \begin{subfigure}[ht]{0.46\textwidth}
         \centering
         \includegraphics[width=\textwidth]{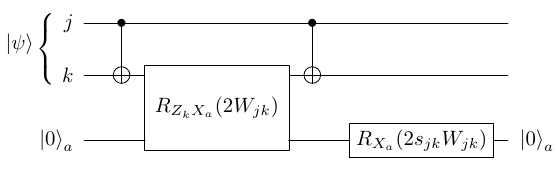}
         \caption{RBM inspired block encoding requires post-selection of the ancillary qubit on $| 0 \rangle$.}
         \label{fig:RBM1}
     \end{subfigure}
     \hfill
     \begin{subfigure}[ht]{0.46\textwidth}
         \centering
         \includegraphics[width=0.8\textwidth]{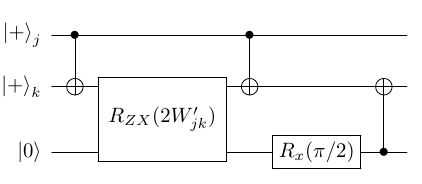}
         \caption{Conditional operation based on the ancilla qubit for the physical initial state $|+\rangle_j\otimes |+\rangle_k$ removes the need for post-selection. }
         \label{fig:RBM2}
     \end{subfigure}
    \caption{Block-encoding-based circuit implementations for the imaginary-time propagator $e^{-\tau w_{jk}Z_jZ_k/2}$ with (a) a generic input state $|\psi\rangle$ and (b) the input state $|+\rangle_j \otimes |+\rangle_k$. Circuit parameters are given in Eq.~\eqref{eq:rbm_para} and Eq.~\eqref{eq:rbm_corr_para}. These block encodings can be applied for a generic Pauli string propagator $e^{- K \prod_i \sigma_i}$.}
    \label{fig:rbm_protocols}
\end{figure*}

\section{Preliminary}
\label{sec:prem}
\subsection{The MaxCut Problem Definition}
In this work, we focus on the MaxCut problem for a given undirected graph $G= (V, \tilde{E})$, where $V= \{1,\dotsc, N\}$ represents the set of vertices, and $\tilde{E} = \{(\langle j,k\rangle, w_{jk})\}$ denotes the set of edges with non-negative weights $w_{jk}\geq 0$ associated with the edge connecting vertices $j$ and $k$. The objective of this problem is to partition vertices of graph $G $ into two disjoint subsets such that the sum of weights of the edges linking the two subsets is maximized. 

To solve the MaxCut problem, one approach is to formulate it as an integer quadratic programming problem, where the goal is to find a solution $\vec{s} \in \{-1, +1\}^{N}$ that maximizes the cost function
\begin{equation}
    C(\vec{s}) = \frac{1}{2}\sum_{\langle j, k\rangle \in \tilde{E}} w_{jk}(1 - s_js_k).
\end{equation}
Here, the solution $\vec{s}$ defines a partition of the graph $G$, with each node $j$ is labeled by $s_j \in\{-1, +1\}$ indicating which of the two subsets it belongs to. Approximation algorithms, such as the Goemans-Williamson algorithm~\cite{goemans1995improved}, are employed to find near-optimal solutions efficiently. The Goemans-Williamson algorithm gives the highest known approximation ratio of $r\approx 0.87856$ among all other classical algorithms for generic graphs ~\cite{goemans1995improved}, while the bound becomes $r\approx 0.9326$ for u3R graphs \cite{halperin2004max}, where $r = C(\vec{s}) / C_{\rm max}$ and $C_{\rm max}$ is the maximum total cost of the cut. 

Quantum computing-based methods can also solve the MaxCut problem, which are designed to search for a quantum state that maximize the quantum cost function $ -\langle H_C \rangle$. The diagonal Hamiltonian $H_C$ for the MaxCut problem on graph $G$ is given by
\begin{equation}
    H_{C} = -\sum_{\langle j, k\rangle \in \tilde{E}}\frac{w_{jk}}{2}(I - Z_jZ_k),
\end{equation}
where $I$ is identity operator for $N$ qubit Hilbert space, $Z_j$ and $Z_k$ are Pauli-Z operators acting on qubits corresponding to vertices $j$ and $k$.  Ideally, we expect the quantum algorithm to return a quantum state $|\psi\rangle$ as a superposition of computational basis $|\vec{z}\rangle$, where $\vec{z}\in \{0, 1\}^{N}$ is a binary string representing a solution of the problem. In this context, each state $|\vec{z}\rangle$ is an eigenvector of $Z_jZ_k$ with eigenvalue $\pm 1$ for any edge $\langle j, k\rangle$, and the quantum cost function $C(\vec{z}) = -\langle \vec{z}|H_C|\vec{z}\rangle$ has a similar structure to the classical cost function $C(\vec{s})$ in the integer quadratic programming problem. 

\subsection{Quantum Approximate Optimization Algorithm}
The Quantum Approximate Optimization Algorithm (QAOA) is a parameterized quantum algorithm designed to tackle general combinatorial optimization problems, including MaxCut. A $p-$level QAOA utilizes two parameterized unitary operators, $U(\gamma) = e^{-i\gamma H_C}$ and $U(\beta) = e^{-i\beta H_B}$, to construct a wavefunction for the graph $G = (V, E)$, 
\begin{equation}
    |\psi( \vec{\beta}, \vec{\gamma} )\rangle = e^{-i\beta_p H_B}e^{-i\gamma_p H_C}\dotsc e^{-i\beta_1 H_B}e^{-i\gamma_1 H_C}|+\rangle^{\otimes N}.
\end{equation}
Here, $H_B = \sum_{j=1}^{N} X_j$ is called the mixing Hamiltonian, where $X_j$ are Pauli-$X$ operators acting on qubit $j$. The expectation value of $H_C$ can be obtained by repeated measurements of the quantum system in the computational basis, 
\begin{equation}
    \langle H_C (\vec{\beta}, \vec{\gamma})\rangle = \langle \psi( \vec{\beta}, \vec{\gamma})| H_C|\psi( \vec{\beta},  \vec{\gamma})\rangle 
    = \sum_{\vec{z}}{\rm Pr}(\vec{z})\langle \vec{z}|H_C|\vec{z}\rangle,
\end{equation}
where $\Pr(\vec{z})$ is the probability of measuring the computational basis state $|\vec{z}\rangle$. 

To find the optimal parameters $(\vec{\beta}, \vec{\gamma})$, one typically starts from an initial guess and then performs numerical optimization to maximize $\langle H_C (\vec{\beta}, \vec{\gamma})\rangle$ using feedback from the quantum computer. This iterative process continues until the maximum cost function value is found. A good choice of initial parameters can significantly reduce the computational cost of optimization. Heuristic strategies have been proposed to optimize $\langle H_C (\vec{\beta}, \vec{\gamma})\rangle$ with respect to parameters $( \vec{\beta}, \vec{\gamma})$ using a good initial guess \cite{zhou_quantum_2020}. Moreover, instead of searching on an instance-by-instance basis, the optimal parameters can be determined in advance for some special cases, including $p\leq8$ QAOA for Sherrington-Kirkpatrick (SK) model~\cite{farhi2022quantum} and $p\leq17$ QAOA for MaxCut on large-girth $D$-regular graphs~\cite{basso_skmodel_2021}. These values have been adopted in some software packages for QAOA simulation such as \texttt{QOKit} from JPMorgan Chase~\cite{qokit}.

\subsection{ITE-BE Method}
Imaginary-time evolution (ITE) is a widely used approach to determine the ground state of a Hamiltonian $H$, by evolving the quantum system  through the imaginary-time propagator
\begin{equation}
    \ket{\psi(\tau)}=e^{-\tau H}\ket{\psi(0)}, \tau >0.
\end{equation}
As long as the initial state $\ket{\psi(0)}$ is not orthogonal to the ground state and $\tau$ is sufficiently large. 

The ITE method can also be applied to classical optimization problems if the classical cost function can be expressed as the expectation value of a problem Hamiltonian. For example, the ITE method can solve the MaxCut problem by finding ground states of Hamiltonian $H_C$, which maximize the MaxCut cost function. In this case, the imaginary-time propagator is given by
\begin{equation}
    e^{-\tau H_C} = e^{\tau  \sum_{\langle j, k\rangle }\frac{w_{jk}}{2}I}e^{-\tau  \sum_{\langle j, k\rangle }\frac{w_{jk}}{2}Z_jZ_k}.
\end{equation}
Here, $e^{\tau \sum_{\langle j, k\rangle}\frac{w_{jk}}{2}I}$ is a global phase factor that can be ignored, and the second term can be exactly decomposed as 
\begin{equation}\label{eq:ite_decompose}
    e^{-\tau  \sum_{\langle j, k\rangle } \frac{w_{jk}}{2}Z_jZ_k} =\prod_{\langle j, k\rangle} e^{-\tau\frac{w_{jk}}{2}Z_jZ_k},
\end{equation}
which does not incur Trotter error because all $Z_jZ_k$ operators in the sum on the left-hand side commute with each other. Therefore, we can choose a sufficiently large time step $\tau$ to ensure convergence. Since each term on the right-hand side is non-unitary, we implement them using a block-encoding-based method proposed in Ref.~\cite{rrapaj_exact_2024}. The circuit block for $e^{-\tau\frac{w_{jk}}{2}Z_jZ_k}$ is shown in Fig.~\ref{fig:RBM1}, and the parameters in the circuit block can be determined analytically by 
\begin{equation}
    \begin{split}\label{eq:rbm_para}
    W_{jk} &= \frac{1}{2}\cos^{-1}\left[ \exp(-|\tau w_{jk}|) \right],\\
    s_{jk} &= {\rm sign}(W_{jk}).
    \end{split}
\end{equation}
In this approach, data will be discarded unless the ancilla qubit is measured to be $|0\rangle$. If the input state to a certain circuit block is $|+\rangle_j \otimes |+\rangle_k$, we can modify the block encoding parameters to deterministically apply the imaginary time propagator, see Fig.~\ref{fig:RBM2}. In particular, the last $\rm CNOT$ gate could ``fix''the data qubit if ancilla is measured to be $|1\rangle$, so no post-selection is needed for this circuit block. We refer readers to Appendix~\ref{app:BE} for further details. For a given undirected graph $G = (V, E)$, the ITE-BE method requires $|V| + 1$ qubits if reusing ancilla qubit for each block in Fig.~\ref{fig:rbm_protocols} and a circuit with $O(|E|)$ depth according to Eq.~\eqref{eq:ite_decompose}.

\section{Protocols and Evaluation}
\label{sec:eval}
\begin{figure*}[ht]
    \centering
    \includegraphics[width=\textwidth]{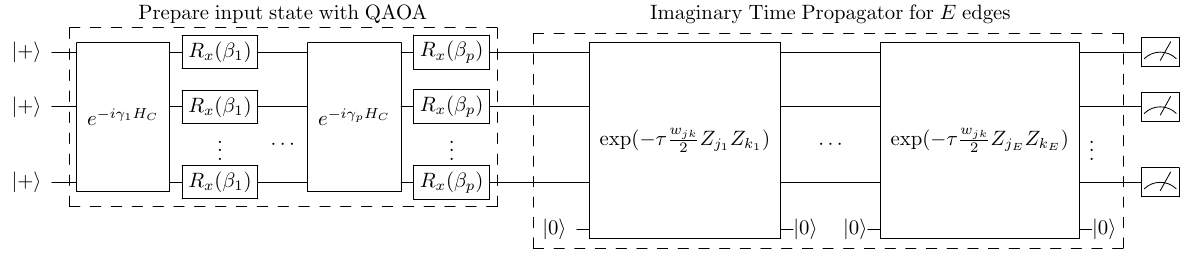}
    \caption{\label{fig:qaoarbm_protocol} The protocol of QAOA+ITE-BE method with a $p-$level QAOA circuit state preparation for ITE-BE rather than using the equal superposition state $|+\rangle^{\otimes n}$.  }
\end{figure*}

In this section, we design two protocols built upon the imaginary-time evolution to solve the MaxCut problem. The first protocol is a purely ITE-BE method starting with initial state $|+\rangle^{\otimes N}$. In this case, we adopt the circuit shown in Fig.~\ref{fig:RBM2} for the first layer, which improves the post-selection rate, and apply the circuit in Fig.~\ref{fig:RBM1} for the subsequent layers. The second protocol starts with the QAOA circuit, followed by ITE-BE, see Fig.~\ref{fig:qaoarbm_protocol}. In essence, this method uses a QAOA circuit to prepare an initial state with a non-minimal overlap with the ground state. Then, the state is evolved through ITE-BE to significantly enhance the overlap. 

Both methods can be applied to general settings. As proof of concept, we numerically evaluate the performance of these protocols on unweighted 3-regular (u3R) graphs with vertex number $6\leq N \leq 12$. We finally report the $99.7\%$ uncertainty intervals for approximation ratio $r$, the probability of obtaining the optimal solution $p_{\rm opt} = N_{\rm opt} / N_{\rm tot}$ and the post-selection rate. Here, $N_{\rm tot}$ is the number of measurements made on data qubits after the post-selection of the ITE-BE method, and $N_{\rm opt}$ is the number of times one of the optimal solutions is obtained. Both metrics estimate the average quality of an algorithm, but $p_{\rm opt}$ is more conservative because it directly reflects the algorithm's ability to find exact solutions. The post-selection rate denotes the success rate of the block encoding procedure, which is the probability in getting all $\ket{0}$ state for the ancilla qubits.

\subsection{ITE-BE Only}\label{sec:itebe_only}
\begin{figure*}[ht]
    \centering
    \includegraphics[width=\textwidth]{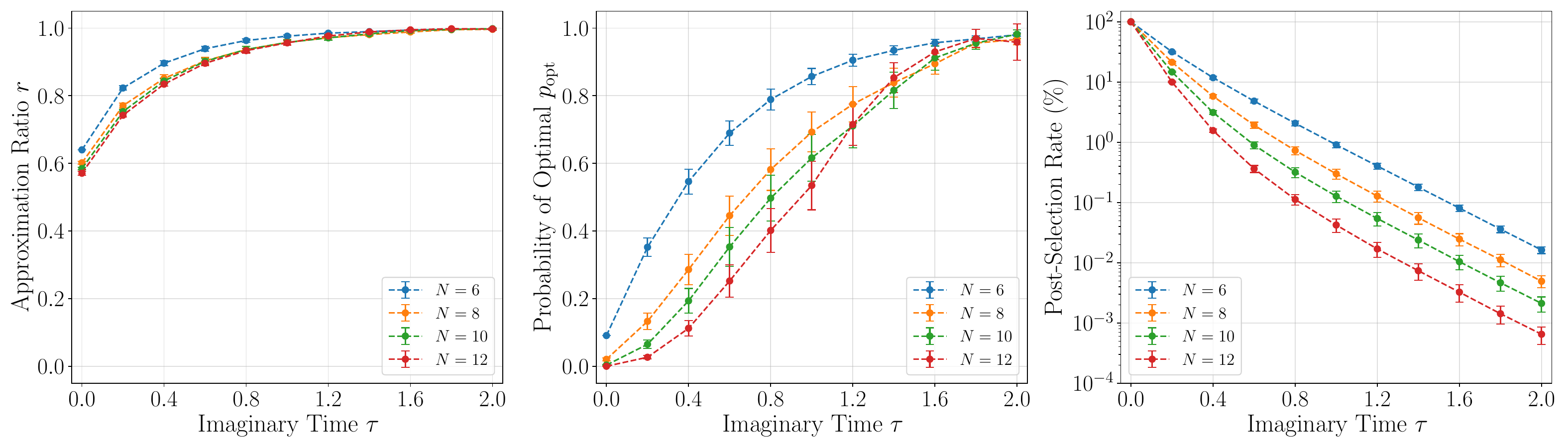}
\caption{\label{fig:newrbm} The averaged approximation ratio $r$, probability of obtaining optimal solution $p_{\rm opt}$, and the post-selection success rate for randomly generated unweighed $3$-regular graphs within three standard deviations of the mean for ITE-BE method. Here, the approximation ratio starts from $0.5$ at $\tau=0$, which is expected according to Section~6.2.1 of Ref.~\cite{prob_and_computing}. This method guarantees convergence to the optimal solution, while the post-selection success rate decays exponentially. The convergence becomes slower and the success rate decreases when the problem size $N$ increases. }
\end{figure*}
Our first method is purely based on the imaginary time evolution. Since all $ Z_jZ_k$ terms commute in the sum on the left-hand side of Eq.~\eqref{eq:ite_decompose}, we can rearrange the order of multiplication on the right-hand side of the same equation and separate them into two parts $V_1$ and $V_2$. For the first part, we define $V_1 = \prod_{\langle j, k\rangle}e^{-\tau\frac{w_{jk}}{2}Z_jZ_k}$ using edges $\langle j, k\rangle$ that form a maximal matching in the graph, i.e., a set of disjoint edges without common vertices that includes as many nodes as possible. All imaginary propagators in $V_1$ can be implemented by the circuit with correction shown in Fig.~\ref{fig:RBM2} when the protocol starts with initial state $|+\rangle^{\otimes N}$. The second part consists of all remaining edges in the graph, and they can be only implemented using the circuit in Fig.~\ref{fig:RBM1} as the input state of each block is no longer $|+\rangle_j\otimes |+\rangle_k$. We have verified that this choice enhances post-selection success. To reduce the overhead of this approach, ancilla qubits for each block can be reused by performing mid-circuit measurements and resetting them to $|0\rangle$. In this case, the current shot terminates and a new shot begins if any of the ancilla qubits in the second part is measured to be $|1\rangle$. All data qubits are measured only after all circuit blocks for the imaginary time propagator $e^{-\tau H_C}$ have been executed. 

We then apply this method to solve the MaxCut problem on 100 randomly u3R graph instances generated by \texttt{networkx} Python package with fixed $N$, where $w_{jk} = 1$ for any $j,k$ in edges $E$ and each node is connected exactly to three other nodes. The imaginary time $\tau$ is varied from $0$ to $2$ which allows us to observe how the results evolve until they are sufficiently close to the ideal solutions. For each $\tau$, we run the ITE-BE circuit with $10^5$ shots for each individual graph and calculate the approximation ratio, the probability of obtaining the optimal solution, and the post-selection success rate using data from successful executions. We repeat this process 10 times to accumulate enough samples and reduce statistical fluctuations. Finally, we exclude graphs with at least one overall failure (i.e., all $10^5$ shots discarded) in 10 repetitions, and report $99.7\%$ uncertainty intervals of above three metrics in Fig.~\ref{fig:newrbm}. All quantum circuits were simulated with the \texttt{AerSimulator} backend provided by the \texttt{Qiskit} Python package~\cite{qiskit2024}. 

Our simulation shows that, as expected, both the approximation ratio $r$ and the probability of obtaining optimal solutions $p_{\rm opt}$ increase monotonically with imaginary time $\tau$ across all graph sizes and eventually saturate near the optimal value of $1.0$ as $\tau$ approaches $ 2$. Meanwhile, the post-selection success rate decreases exponentially with $\tau$, which is consistent with the prediction in Ref.~\cite{rrapaj_exact_2024}. Such  behavior is typical of repeated applications of block encoding due to the post selection probability being less than 1. If one chooses to employ a separate auxiliary qubit per Pauli string in the Hamiltonian, and post select at the end of all propagator applications, then one can resort to amplitude amplification~\cite{Bassard:1997} to improve the post-selection rate. In addition to the increased cost in terms of the number auxiliary qubits, this additional step would also increase the overall circuit depth. We note that smaller graphs converge more quickly to the ground state and exhibit higher post-selection success rates compared to larger graphs due to fewer terms in the Hamiltonian. 

Further analysis shows that for bipartite graph, the ITE-BE method shows a faster convergence to the ground state and higher success rate compared to non-bipartite graphs of the same size. As a result, we do not consider bipartite graphs in Fig.~\ref{fig:newrbm} and provide a detailed investigation of the ITE-BE method on MaxCut problem for them in Appendix~\ref{app:outliers}.

\subsection{QAOA + ITE-BE}
\begin{figure*}
    \centering
    \includegraphics[width=\textwidth]{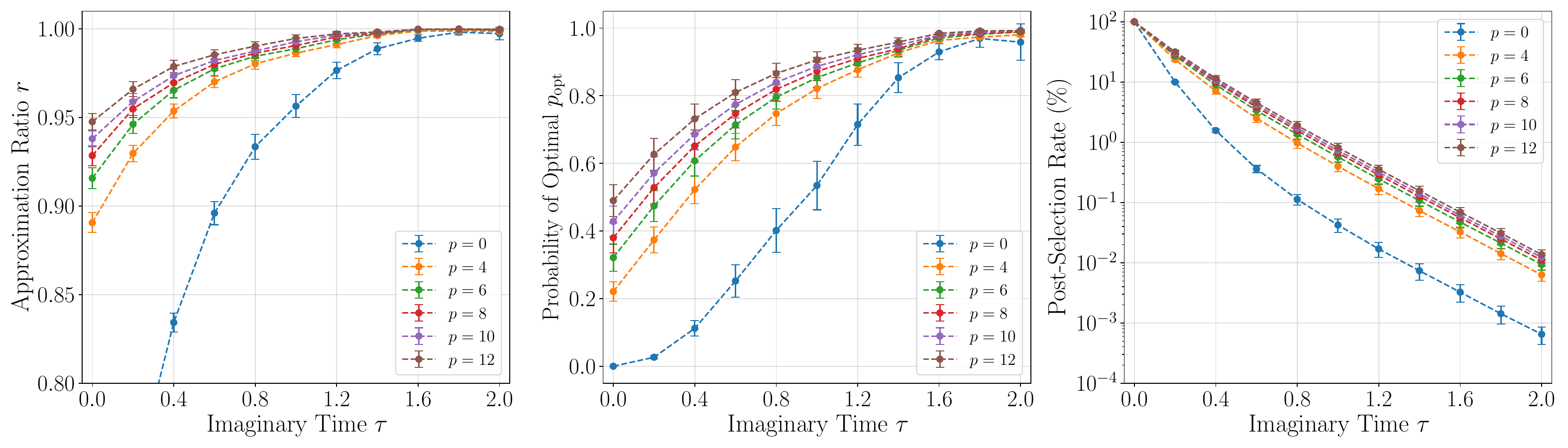}
\caption{\label{fig:qaoarbm} The averaged approximation ratio $r$, probability of obtaining optimal solution $p_{\rm opt}$, and the post-selection success rate within three standard deviations of the mean for the QAOA+ITE-BE method. Here we use $p-$level QAOA circuit for state preparation, where $p = 4,6,8,10,12$. The QAOA input state leads to faster convergence of ITE-BE and a post-selection rate that is one order of magnitude higher than using an equal superposition state to solve same graph for $p=12$. As comparison, we show the results from purely ITE-BE method labeled as $p=0$ in above figures, which indicates that the ITE-BE has faster convergence to the optimal solution and nearly $10$ times higher post-selection rate when using $p$-level QAOA input state. }
\end{figure*}

Our second method combines the QAOA with the imaginary time evolution (see Fig.~\ref{fig:qaoarbm_protocol}). In the first part of the protocol, we directly use near-optimal values $(\vec \beta, \vec \gamma)$ presented in Ref.~\cite{basso_skmodel_2021} without any further training of QAOA to prepare an input state for the ITE-BE part. The QAOA output state has a higher overlap with the ground state than the standard equal superposition state $|+\rangle^{\otimes n}$. As such we expect this protocol to improve the post-selection success rate in the subsequent ITE-BE part. In the second part, we apply the ITE-BE method to further search for the optimal solution, where each imaginary time propagator for edge $\langle j,k\rangle$ is implemented by the circuit block in Fig.~\ref{fig:RBM1}. To implement the overall protocol for a given undirected graph $G = (V, E)$, we could reuse ancilla qubits for each circuit block in the ITE-BE part such that only $|V| + 1$ qubits is required. The data qubits are measured only after all circuit blocks have been executed. Also, the circuit depth of $e^{-i\gamma_k H_C}$ in state preparation is $O(|E|)$, so the circuit for overall protocol has $O(p|E| + |E|)$ depth.

In order to evaluate its performance, we apply this hybrid approach to solve the MaxCut problem on the same set of randomly generated unweighted 3-regular(u3R) graph instances as in Sec.~\ref{sec:itebe_only},  following the same settings.  We report the $99.7\%$ averaged uncertainty interval for the averaged approximation ratio $r$, probability of optimal $p_{\rm opt}$ and post-selection success rate as a function of imaginary time $\tau$ for graphs with $N=12$ and different QAOA levels in Fig.~\ref{fig:qaoarbm}.

Similar to the pure ITE-BE results, both $r$ and $p_{\rm opt}$ increase monotonically with time $\tau$, and the final results always converge to the ground state at $\tau\sim 2$ regardless of the QAOA level. Also, the post-selection success rate decays exponentially with $\tau$,  but this trend is independent on the QAOA level. These observations suggest that a shallow QAOA+ITE-BE circuit with a lower $p$ and a large $\tau$ is sufficient to obtain the optimal solution. Additional results for graphs with $N = 6,8,10$ lead to the same conclusion, and the data can be found in Appendix~\ref{app:plots}. For bipartite graphs, this method demonstrates rapid convergence and high success rate which eventually reach a plateau. We do not consider bipartite graphs in Fig.~\ref{fig:qaoarbm}, but study the performance on Appendix~\ref{app:outliers}.

It is worth noting that at $\tau = 0$, the imaginary time propagator does not act on the input state. Thus, the values of both $r$ and $p_{\rm opt}$ at $\tau = 0$ reflect the performance of $p-$level QAOA with near-optimal parameter values from Ref.~\cite{basso_skmodel_2021}. As shown in Fig.~\ref{fig:qaoarbm}, the QAOA output state achieves a higher approximation ratio but a lower probability of obtaining optimal solution. This behavior indicates that the QAOA output state is a superposition of the ground state and multiple excited state, $|\psi\rangle = \sum_{j}c_j|\vec{z}_j\rangle$. The presence of excited states lead to a lower $p_{\rm opt}$, but their total contribution to the cost function $\langle H_C\rangle$ is significantly smaller than that of the ground state, which results in a high approximation ratio. We can conclude that parameter values in Ref.~\cite{basso_skmodel_2021} are tuned to maximize $r$ rather than $p_{\rm opt}$, and the ITE-BE layer significantly enhance performance of QAOA to achieve a higher $p_{\rm opt}$ while also improving $r$. 

Another interesting comparison can be made on the post-selection rate as function of the initial state. Not only does the ITE-BE method improve the results of QAOA, but also the intitial state provided by QAOA increases the post-selection rate with respect to the  $|+\rangle^{\otimes n}$. Indeed, for $\tau=2$ and $N=12$ the post-selection rate increases by about an order of magnitude. 

\section{Conclusion and Outlook}\label{sec:dis}
In this paper, we propose two quantum algorithms based on the block-encoding implementation of the imaginary time evolution propagator (ITE-BE) and evaluate their performance on the MaxCut problem. We conduct numerical simulations on randomly generated unweighted $3-$regular graphs with vertex number from $N = 6$ to $N = 12$. Using circuits with $N$ qubits and $O(|E|)$ depth, our results show that both ITE-BE based methods converge to the ground state when the imaginary time $\tau$ is sufficiently large. This performance surpasses that of the QAOA algorithm. In comparison to classical MaxCut solvers, current semi-definite programming approaches can solve dense graph instances (density up to $0.9$) for up to $N=100$ nodes (\texttt{BiqMac}~\cite{biqmac}) or $N=180$ nodes (\texttt{BiqBin}~\cite{biqbin}). Moreover, modern linear programming solver can handle sparse graphs with hundreds or thousands of nodes~\cite{Rehfeldt2023}, and even more than $N=10{,}000$ nodes (\texttt{McSparse}~\cite{McSparse}). Thus, demonstrating quantum advantage would require access to more than $100$ qubits. However, the guaranteed convergence of the ITE-BE based methods still indicates their promising potential for solving MaxCut problems on large graphs. 

The success of QAOA+ITE-BE method emphasizes the importance of concatenating quantum algorithms in order to boost overall performance and overcome individual limitations.  When preparing the initial state with a $p-$level QAOA for the ITE-BE layer, our results show a faster convergence of ITE-BE to the optimal solution and a post-selection rate that is an order of magnitude higher than using $|+\rangle^{\otimes n}$ as initial state. The result of QAOA+ITE-BE also reveals a trade-off between the QAOA level $p$ and the imaginary time $\tau$, namely, one can choose a lower QAOA level with a large $\tau$ to ensure convergence using a shallow circuit, or choose higher QAOA level to have quicker convergence with higher post-selection rate. Moreover, integrating QAOA with the ITE-BE method also benefits QAOA, as it avoids the need for additional training of QAOA with near-optimal parameters to achieve better accuracy. Instead, the subsequent ITE-BE layer enhances the result with only $O(|E|)$ circuit depth, where $|E|$ is the number of edges in the graph. This improvement is particularly significant when solving MaxCut on large graphs.

Our work focuses mainly on the MaxCut problem,  but the principle of both the ITE-BE and QAOA+ITE-BE methods can be applied to other combinatorial optimization problems such as the Traveling Salesman problem or the Knapsack problem. The practical application of these methods is limited by the exponential decay in post-selection success rates with increasing imaginary time $\tau$ and graph size $N$. Therefore, developing methods to improve the post-selection success rate or enhance the convergence speed is a crucial research direction we leave for exploration in future work. By solving this challenge with advanced algorithm designs, we will be able to realize the full potential of imaginary time evolution on optimization problems and achieve practical quantum advantage.

\section{Acknowledgment}
This research was supported by the U.S. Department of Energy (DOE) under Contract No. DE-AC02-05CH11231, through the National Energy Research Scientific Computing Center (NERSC), an Office of Science User Facility located at Lawrence Berkeley National Laboratory.
A.F. acknowledges funding by the U.S. Department of Energy, Office of Science, Office of Advanced Scientific Computing Research Quantum Testbed Program under contract DE-AC02-05CH11231. The authors acknowledge the Center for Advanced Research Computing (CARC) at the University of Southern California for providing computing resources that have contributed to the research results reported within this publication. URL: https://carc.usc.edu.

\bibliography{ref}

\appendix 
\clearpage

\section{ITE Parameters}\label{app:BE}
We provide here the general form of the ansatz for a single qubit (the same analysis applies to n-qubit Pauli string),
\begin{equation}
e^{-\tau K \sigma_r}=\mathcal{N} \langle 0_a|e^{-i\left(c \sigma^x_a +W\sigma_r \sigma^x_a\right)}| 0_a \rangle,
\label{eq:rbm_ansatz}
\end{equation}
where $W$ is the coupling between ancilla qubit $a$  and the target qubit r.  
For a generic initial state, $|\Psi\rangle = \alpha |0\rangle + \beta |1\rangle$ with $|\beta|^2=1-|\alpha|^2$, the ratio of the probability to measure $| 0_a \rangle$ versus $| 1_a \rangle$ is,
\begin{equation}
    \frac{P(0_a)}{P(1_a)}= \left| \frac{\alpha  \cos (c+W)+\beta  \cos (c-W)}{\alpha  \sin (c+W)+\beta  \sin (c-W)}\right|^2.
\end{equation}

In~\cite{rrapaj_exact_2024}, the values of the parameters are,
\begin{equation}
\begin{split}
\mathcal{N}&=\exp(| \tau K|)/2,\\
 W&=\frac{1}{2} \cos^{-1}\left[\exp(-2|\tau K|)\right],\\ 
c &= W s,\ s = \mathrm{sign}(K).
\end{split}
\label{Rbm params}
\end{equation}
When post-selection fails, one must repeat the operation. However, for special cases, one can correct instead of re-starting computations. For instance, instead of the parameters in the \eqref{Rbm params}, we can use the following parameters, 
\begin{equation}
\begin{split}
\mathcal{N}&=\frac{\sqrt{e^{4 \tau K}+1}}{2 \sqrt{e^{2 \tau K}}},\\
 W&=\tan ^{-1}\left(e^{2 \tau K}\right) - \pi/4,\\ 
 c &=\pi/4.
\end{split}\label{eq:rbm_corr_para}
\end{equation}
The state post operation is,
\begin{equation}
\begin{split}
    U(|\Psi\rangle \otimes |0_a\rangle) =& \frac{1}{\sqrt{e^{4 K \tau }+1}} (\alpha |0\rangle + \beta e^{2 K \tau } |1\rangle) \otimes |0_a\rangle \\
    -& \frac{i e^{2 K \tau }}{\sqrt{e^{4 K \tau }+1}}(\alpha |0\rangle + \beta e^{-2 K \tau } |1\rangle) \otimes |1_a\rangle.
\end{split}
\end{equation}
Thus, when the ancilla is measure in the $|0_a\rangle$ state, the desired operation is performed in the system qubit, and when the ancilla is measure in the $|1_a\rangle$ state, the opposite operation is performed. If $\alpha = \beta$, we can flip the amplitudes with a $CX$ gate, effectively correcting post-selection, and no measurement is required. The ancilla can be reset for re-use.

\section{The MaxCut Solution of Bipartite Graphs}\label{app:outliers}
\begin{figure*}[ht]
    \centering
    \includegraphics[width=\textwidth]{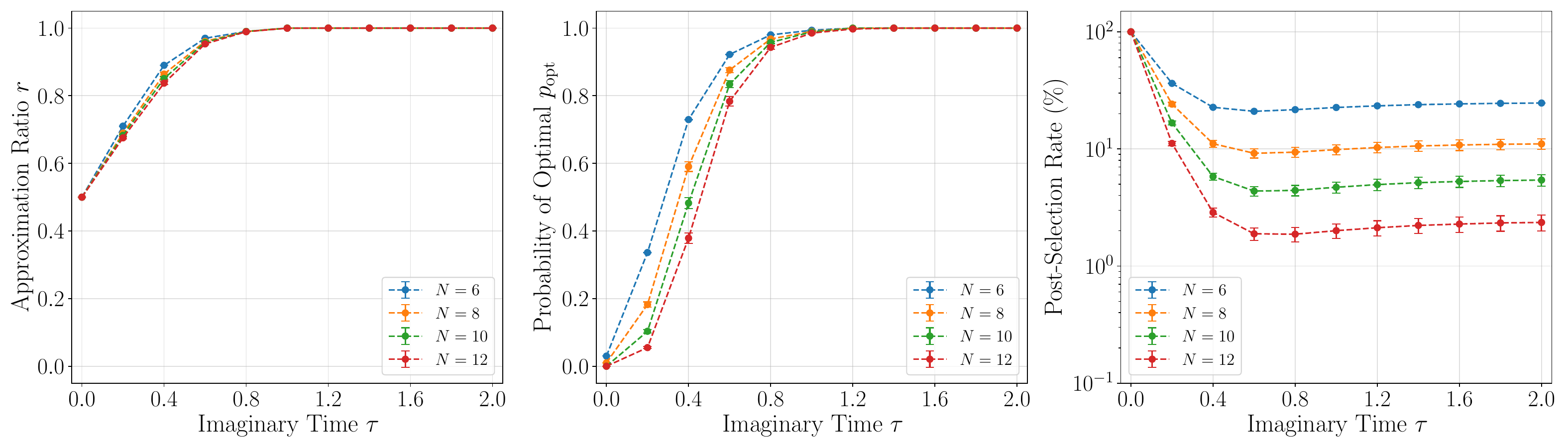}
    \caption{\label{fig:newrbm_bipartite} The averaged performance of pure ITE-BE method on 40 randomly generated bipartite $3-$regular graphs for $N = 6, 8, 10$ and $12$. The ITE-BE method converges faster than non-bipartite graphs and the success rate reaches a plateau after imaginary time $\tau \sim 0.6$. }
\end{figure*}

\begin{figure*}
    \centering
    \includegraphics[width=\textwidth]{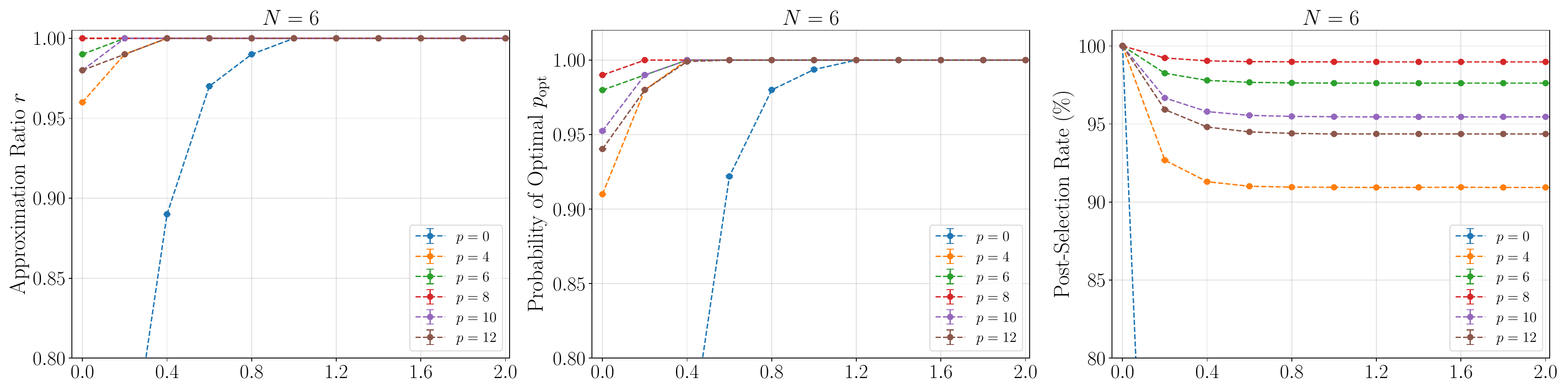}
    \includegraphics[width=\textwidth]{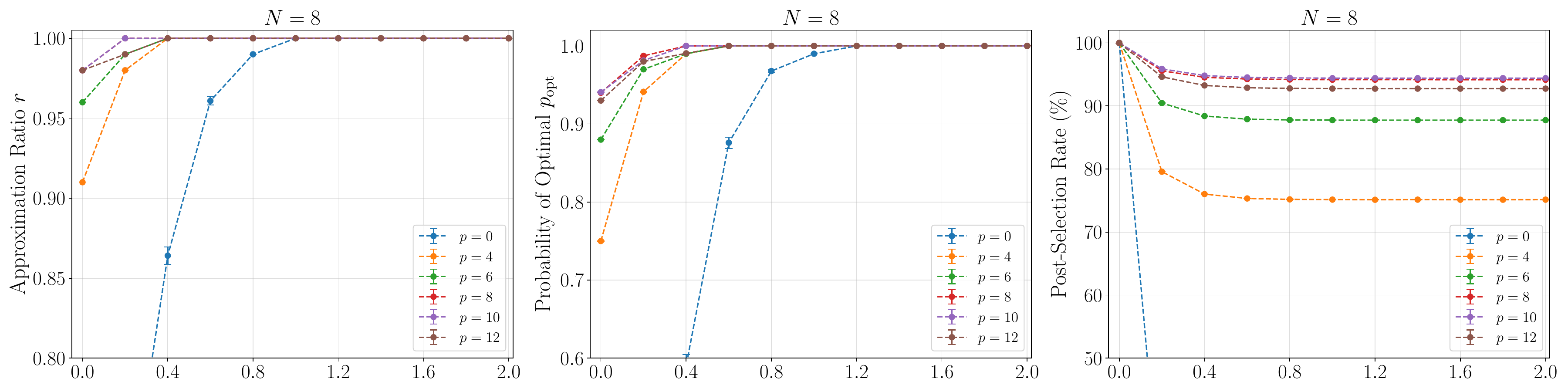}
    \includegraphics[width=\textwidth]{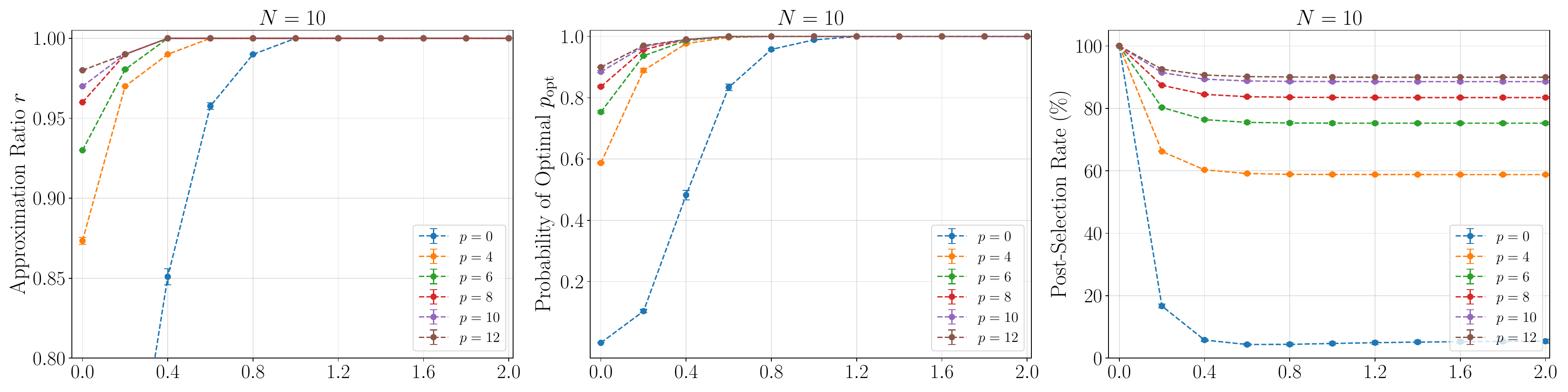}
    \includegraphics[width=\textwidth]{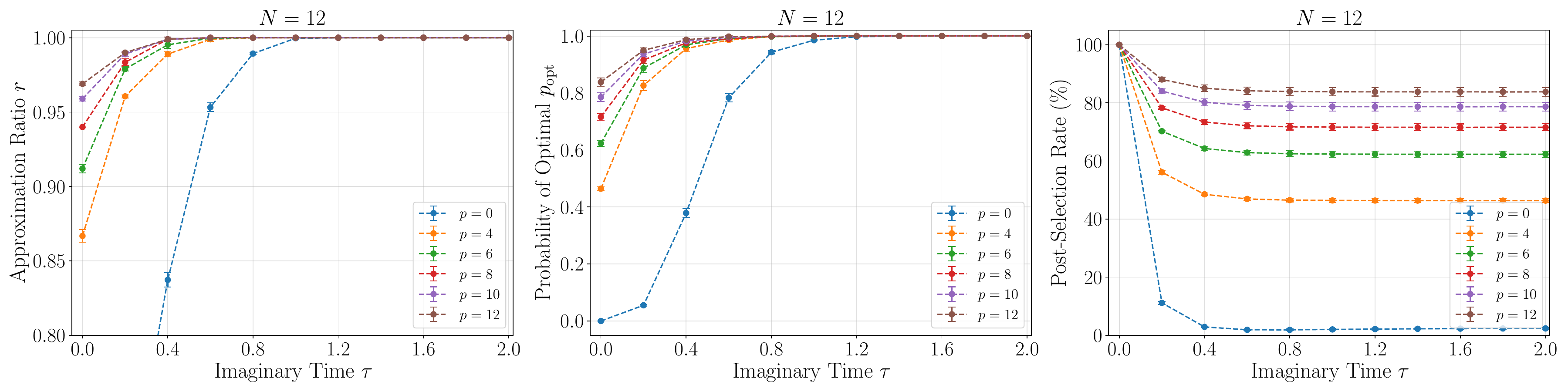}
\caption{\label{fig:qaoarbm_bipartite} The averaged performance of QAOA+ITE-BE method on 40 randomly generated bipartite $3-$regular graphs for $N = 6, 8, 10$ and $12$. In this case, the QAOA initial state is very close to the optimal solution so the ITE-BE method have only smaller improvement. The success rate reaches a plateau after imaginary time $\tau \sim 0.4$. The $p=0$ curves give the results from purely ITE-BE method on same bipartite graphs. }
\end{figure*}

During the process of data analysis, we observe that both ITE-BE methods converge more quickly for some graphs than for others in our sample. Moreover, these graphs have higher post-selection success rate compared to other graph with same size, and the success rate reaches a plateau after a certain imaginary time $\tau$. To understand the conditions under which the ITE-BE method converges more quickly, we compute the spectra of both the adjacency matrices and Hamiltonians of these special graphs. We found that the minimum eigenvalue of the adjacency matrices of these graphs is $-3$, which is a characteristic feature of unweighted bipartite $3-$regular graphs. In addition, the minimum eigenvalue of their Hamiltonian $H_C$ is given by $(\frac{3N}{2}-1)/2$, where $\frac{3N}{2}$ is the total number of edges in a $3-$regular graph. This implies that the MaxCut solution is a partition where all edges in the graph connect the two parts of the partition. These observations conclusively identify the special graphs as bipartite $3-$regular graphs. 

To verify whether the ITE-BE method converges quickly on other bipartite $3$-regular graphs, we randomly generate 40 instances of bipartite graphs with $6\leq N \leq 12$ nodes, and solve the MaxCut problem using both ITE-BE and QAOA + ITE-BE methods. The numerical simulation setting and data analysis process are consistent with those described in the main text. The results are shown in Fig.~\ref{fig:newrbm_bipartite} and Fig.~\ref{fig:qaoarbm_bipartite}. 

It is worth noting that any bipartite graphs with non-negative weights and only one connected component, the solution to the MaxCut problem is unique up to a ``flipping'' of all vertices and corresponds to the bipartition of the graph. This is because, in a bipartite graph, all edges connect vertices from one part to the other, so the total weight of the edges crossing the bipartition is maximized and cannot be increased further.  In this case, the MaxCut solution can be found using classical algorithms such as breadth-first search (BFS) or depth-first search (DFS) with linear time complexity $O(V+E)$. 

\begin{figure*}
    \centering
    \includegraphics[width=\textwidth]{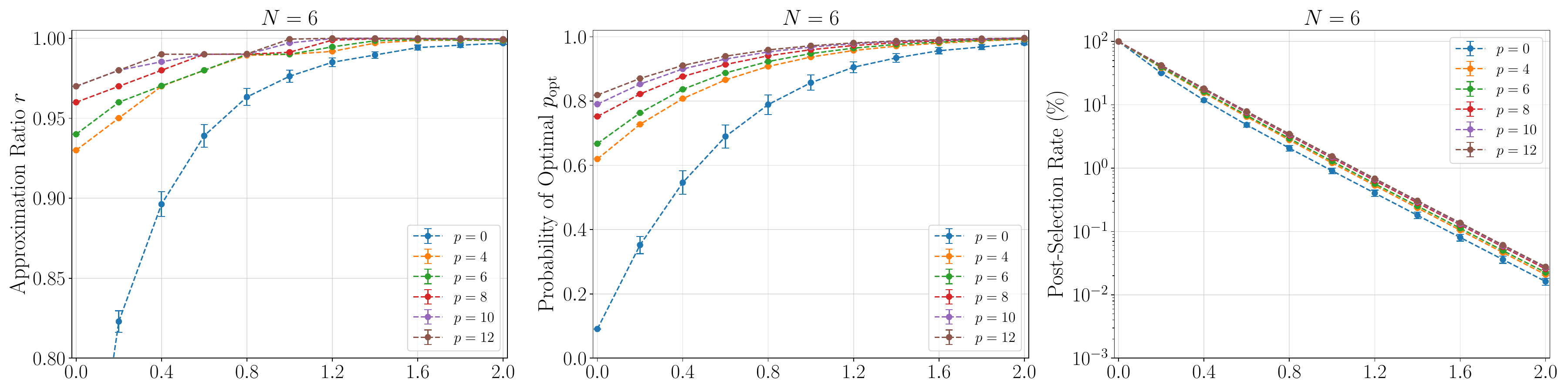}
    \includegraphics[width=\textwidth]{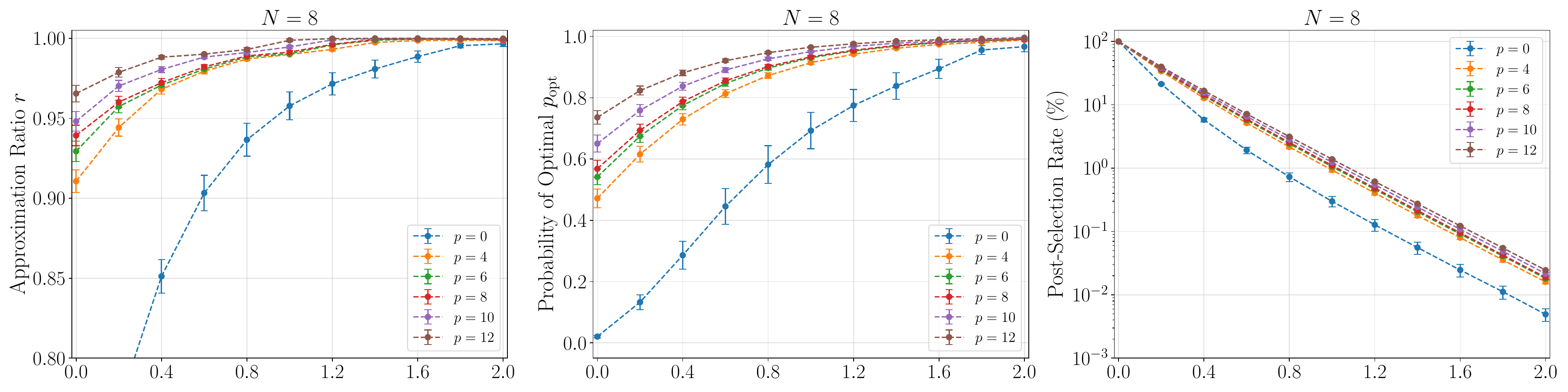}
    \includegraphics[width=\textwidth]{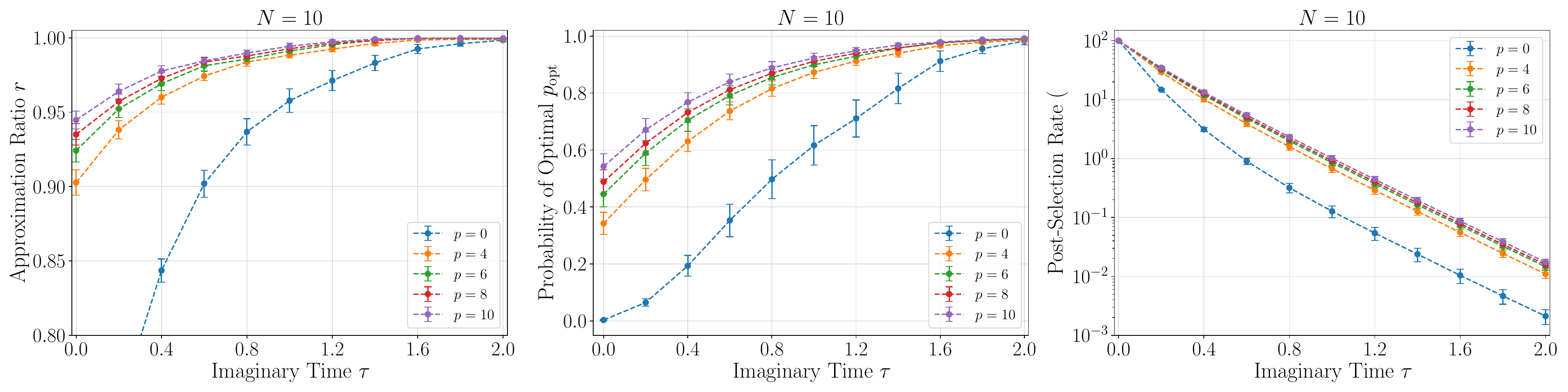}
\caption{\label{fig:qaoarbm_extra} The averaged performance of QAOA+ITE-BE method on 100 randomly generated u3R graphs for $N = 6,8,10$. The ITE-BE layer further enhances the performance of QAOA especially for large graphs. We also show results obtained from purely ITE-BE method ($p=0$) on same u3R graph instances as a comparison. }
\end{figure*}

\section{QAOA+ITE-BE Results for Graphs with 6, 8, 10 Nodes}\label{app:plots}
In this appendix, we show the results of the QAOA+ITE-BE method on u3R graph instances with $N = 6,8$ and $10$ in Fig.~\ref{fig:qaoarbm_extra}. Due to the optimal parameters for QAOA, at $\tau=0$, the approximation ratio is already above $0.9$ and further improves with ITE-BE. The $p_{\rm opt}$ at $\tau=0$, on the other hand, is smaller in value and decreases significantly as the system become larger, compared to approximation ratio, for the optimal QAOA. The application of the ITE-BE layer increases the overlap with the ground state, as expected. 

\end{document}